\documentclass[12pt,preprint]{aastex}
\shorttitle{Disk Around Mira~B}
\shortauthors{Ireland et al.}
\begin{document}

\title{Born-Again Protoplanetary Disk Around Mira~B}

\author{Ireland, M.J.}
\affil{Planetary Science, California Institute of
  Technology, Pasadena, CA 91125}
\email{mireland@gps.caltech.edu}
\and
\author{Monnier, J.D.}
\affil{University of Michigan, Ann Arbor, MI 48109}
\email{monnier@umich.edu}
\and
\author{Tuthill, P.G.}
\affil{School of Physics, University of Sydney, NSW 2006, Australia}
\email{gekko@physics.usyd.edu.au}
\and
\author{Cohen, R.W.}
\affil{W.M.Keck Observatory, Kamuela, HI 96743}
\and
\author{De Buizer, J.M.}
\affil{Gemini Observatory, 950 N. Cherry Ave, Tucson, AZ 85719}
\and
\author{Packham, C.}
\affil{University of Florida, Gainsville, FL 32611}
\and
\author{Ciardi, D.}
\affil{Michelson Science Center,Caltech, Pasadena CA 91125, USA}
\and
\author{Hayward, T.}
\affil{Gemini Observatory, Casilla 603, La Serena, Chile}
\and
\author{Lloyd, J.P.}
\affil{Cornell University, Ithaca, NY 14853, USA}

\begin{abstract}
 The Mira~AB system is a nearby ($\sim$107\,pc) example of a wind
 accreting binary star system. In this class of
 system, the wind from a mass-losing red giant star (Mira~A) is
 accreted onto a companion (Mira~B), as indicated by an accretion
 shock signature in spectra at ultraviolet and X-ray
 wavelengths. Using novel imaging techniques, we report the detection
 of emission at mid-infrared wavelengths between 9.7 and 18.3\,$\mu$m 
 from the vicinity of Mira~B but with a peak at a radial position
 about 10 AU closer to the primary Mira~A.  We interpret the
 mid-infrared emission as the edge of an optically-thick accretion disk
 heated by Mira~A. The discovery of this new class of accretion disk
 fed by M-giant mass loss implies a potential population of young
 planetary systems in white-dwarf binaries which has been little
 explored, despite being relatively common in the solar neighborhood.
\end{abstract}

\section{Introduction}

Mira (o Ceti) is the prototype of the class of Mira-type variable
stars, characterized observationally by large visual photometric
pulsation amplitudes and regular periods between 100 and 1000
days. Mira-type variables represent one of the late stages of
evolution for 1-2 solar mass stars, with luminosities of several
thousand times solar and mass loss rates of order $10^{-6}$ solar
masses per year.  The winds from Mira-type variables are relatively
slow (typically $<$\,10 km\,s$^{-1}$), which means that the mass lost in these
winds can be readily accreted onto any stellar companions. It is known
that 57\% of the progenitors of Mira-type variables, G and F dwarfs,
have a companion with a mass more than a tenth of the primary mass
\citep{Duquennoy91}. However, most of the companions around Mira-type
primaries remain undetected because of the overwhelming flux of the
Mira-type primary and the difficulty in making precise radial velocity
measurements due to the stellar pulsation. The most reliable ways to 
detect companions are through the signs of wind accretion: either a
short-wavelength (e.g. ultraviolet) excess or rapid photometric
variability \citep{Sokoloski01}.


Mira~B was first discovered visually in 1923, when it was at a
separation of 0.9'' \citep{Aitken23}, and has since been imaged
in X-rays, ultraviolet and the radio  
\citep{Karovska97,Karovska05,Matthews06} . Observations of the Mira~AB
system with the Infrared Spatial Interferometer also
offered a tentative detection of flux from near the position of Mira~B at 
11.17\,$\mu$m, interpreted as a clump of dust
\citep{Lopez97}. Previous attempts at deconvolution of mid-infrared 
images \cite{Marengo01} showed a coma-like elongation in the direction
of Mira~B and a feature at $\sim$twice the Mira~AB separation, but
quantitative interpretation of these data was difficult due to the
difficulties in de-convolving seeing-limited images.
 
The nature of Mira~B has been and remains
controversial, due to the radiation originating from Mira~B being
dominated by accretion. According to the arguments originally
made by \citet{Jura84} and reinforced by \citet{Kastner04}, Mira
B must be a low-mass main sequence star, as the X-ray and total luminosity of
Mira~B is far too weak for the companion to be a white dwarf. This
argument relies on measurements of the Mira~B wind, reasonable values
for the true (i.e. non projected) separation of Mira~AB and the Bondi-Hoyle
accretion rate being correct within a factor of 10.

This paper presents novel non-redundant interferometry observations of Mira~AB, 
enabling robust detection of mid-infrared flux from the vicinity of Mira~B. 
These and other traditional imaging observations are presented in detail in 
Section~\ref{sectObservations}. An examination of possible models for
this emission is presented in Section~\ref{sectModelling} and the
implications of the preferred (disk) model presented in Section~\ref{sectDiscussion}.

\section{Observations}
\label{sectObservations}
In the mid-infrared, observations on a 10\,m class of telescope can be
nearly diffraction limited in good seeing conditions. With a 15\,cm
atmospheric coherence length at 550 nm, a 10 m aperture is two
coherence lengths wide at 10\,$\mu$m. In this regime, the image of a
point source usually has a single bright core, but deconvolution is difficult
due to changes in seeing and changes in low order aberrations with changing
elevations. 

We first overcame these difficulties by individually re-pointing and
re-phasing 4 sets of 6 segments out of the 36 separate segments that
made up the primary mirror of the Keck I telescope. As part of an
observing program designed to resolve the dusty winds from Mira
variables, these ``segment tilting'' observations of $o$~Ceti were made
at 8, 9.9, 10.7 and 12.5\,$\mu$m. These four sets of
segments formed 4 interferograms on the Long-Wavelength Spectrometer
(LWS) camera, each of which gave 15 visibility amplitude and 20 closure
phase (10 independent) measurements on the source. This technique is
described in \citet{Monnier04b} and \citet{Weiner06}.  At
10.7\,$\mu$m, data was also taken in a configuration
with 3 sets of 7 segments with non-redundant spacings. Calibration
observations were made of $\alpha$~Cet and $\alpha$~Tau to calibrate
visibility amplitude and subtract out any instrumental closure-phase.

\begin{figure}
 \plotone{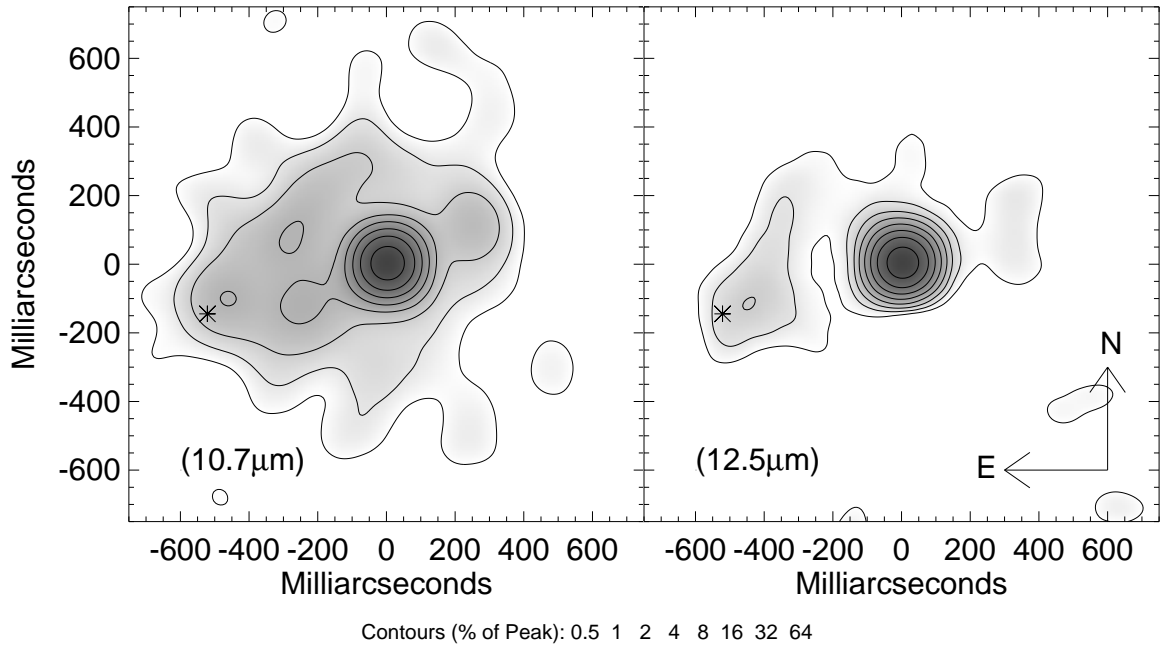}
 \caption{Images of the Mira~AB system reconstructed from
 segment-tilting data taken with the Long Wavelength Spectrometer on
 the Keck I telescope on 2004 August 30. The asterisk in each
 image represents the ultraviolet position of Mira~B extrapolated from
 HST data.} 
\label{figKeckIms}
\end{figure}

The {\tt MACIM} algorithm: a Monte-Carlo technique for imaging from
closure-phase and squared visibility \citep{Ireland06b} was used to
create images at all wavelengths. As part of this imaging process, we
incorporated prior knowledge that a set fraction of the flux comes from a
compact central source (i.e. the central star Mira~A) with a 50\,mas diameter. This size comes from
the longer-baseline observations of \citet{Weiner03}. We estimated the
fraction of light from this 50\,mas source by assuming simple
black-body emission, based on the source containing 75\% of the 8\,$\mu$m flux. 
This step had little effect on flux near the
location of Mira~B, but affected the deconvolution of the central
source and dust shell, due to the documented tendency for techniques
in the Maximum-Entropy family to 
spread flux over the image at the expense of the compact central
source. Unambiguous detections of flux near the location of Mira~B
were made at 12.5 and 10.7 microns, with a lower signal-to-noise
detection at 9.9 microns. The 10.7 and 12.5 micron images are shown in
Figure~\ref{figKeckIms}. 

The separation and position angle of Mira~B with respect to Mira~A was
obtained for the 10.7 and 12.5\,$\mu$m data by searching for a peak in
the difference image between the reconstructed image and the
azimuthally-averaged reconstructed image with Mira~B masked-out. Flux
measurements of Mira~B were obtained from aperture-photometry on this differenced
image. We also attempted to simultaneously 
reconstruct a dust shell image and perform a binary fit to the
Fourier data using {\tt MACIM}, obtaining consistent results and an
independent estimation of errors on the binary parameters directly
from {\tt MACIM}'s Markov Chain. The average separation and position
angle from these fits is $0.48\pm0.01$\arcsec and $104.1\pm1.1$ 
degrees. We independently calibrated the camera plate scale and
orientation by observing binaries XY Per and $\alpha$~Her,
giving results in agreement with previous calibrations
\citep{McCabe06}. Plate scale uncertainties are included in the above
error estimates. 

To obtain comparable astrometry for the ultraviolet position of Mira~B,
we extrapolate based on a linear fit between \citet{Karovska97} and the 
2004 February HST STIS O8WY02040 data sets.
Projected to JD\,2453248, we find a separation of $0.541\pm0.003$'' at
$105.5\pm0.3$ degrees. In 
comparing the ultraviolet and mid-infrared detections, the
position angles are found to be consistent. However, the difference in separation
is significant, so the thermal IR and UV
emission from Mira~B must arise from different physical regions.

To confirm this result, we obtained additional data using
an aperture-mask in the T-ReCS camera on Gemini
South (Program ID GS-2006B-Q-15). The mask is installed in the cold
Lyot stop wheel of T-ReCS, decreasing the background significantly
from the Keck experiment, where the full telescope pupil contributed
to the background in all interferograms. This mask has 7 holes in a 
non-redundant configuration, amounting to 21 simultaneous visibility
amplitudes and 35 closure-phases (15 independent). Baseline lengths
projected on the primary mirror range from 1.78 to 6.43\,m, and sky
rotation is used to increase the Fourier coverage while maintaining
the same relationship between positions on the mask and positions on
the primary mirror. The masking observations of Mira~AB used
$\tau_{04}$~Eri, $\alpha$~Ceti and $\alpha$~Tau as interferometric calibrators.

\begin{figure}
 \plotone{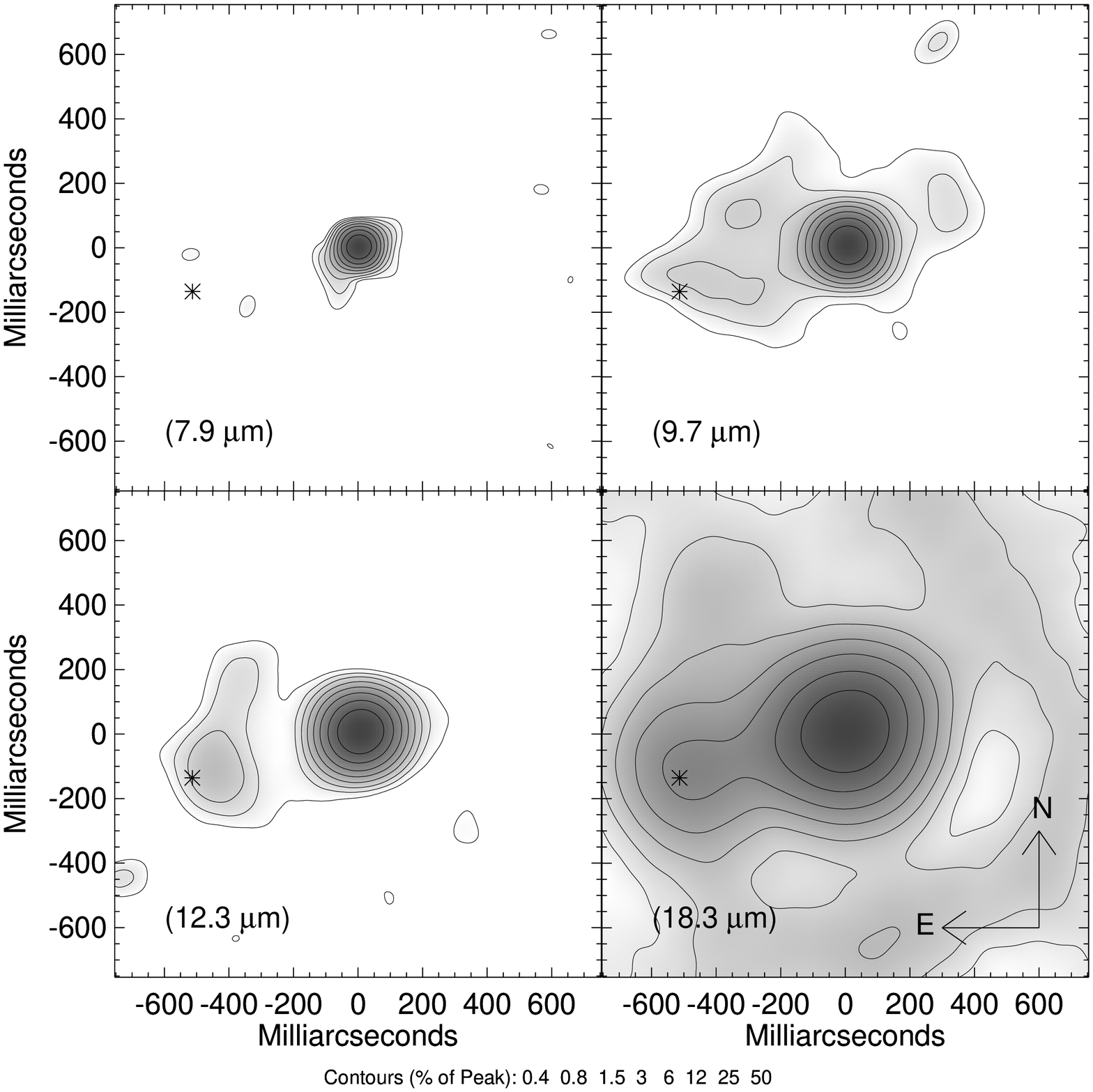}
 \caption{Images of the Mira~AB system reconstructed from 
 aperture-masking data taken with the T-ReCS camera on Gemini South,
 on 2006 November 14 and 16. The asterisk in each
 image represents the ultraviolet position of Mira~B extrapolated from
 HST data.}
\label{figGeminiIms}
\end{figure}

These data were obtained in November 2006 at 7.7\,$\mu$m, 
9.7\,$\mu$m, 12.3\,$\mu$m and 18.3\,$\mu$m. Due to a greater
instrumental stability, these data were of better quality than the Keck
data, although the resolution was slightly less and as Mira~A was near 
minimum, the contrast between Mira~A and the dust shell was
lower. Images reconstructed using {\tt MACIM} are shown in
Figure~\ref{figGeminiIms} - again showing a clear feature in the vicinity of
Mira~B in three out of four filters. Astrometry and flux estimates were
obtained in a similar way to the Keck data, with fits summarized in
Table~\ref{tabBinaryFits}. Flux ratios and astrometry were consistent
between the Keck and Gemini data.


In addition to these mid-infrared aperture-masking observations, we
analyzed earlier Keck/LWS observations of $o$~Ceti from September 1999 and
August 2002 made without the use of a mask or segment-tilting.
Rotating the images 180 degrees and subtracting enabled most
of the primary PSF to be removed. 
Again, a bright feature in the vicinity of Mira~B was detected at
a high level of confidence. 
The contrast ratios obtained from
these data were 0.102$\pm$0.022 in September 1999 (JD 2451423.6) and
$0.11\pm0.03$ in August 2002 (JD 2452506.6). The errors were estimated
by applying the same analysis method to 6 calibrator star
observations. 

Other observations at wavelengths between 0.97 microns (HST, NICMOS data set
N4RK02OTQ) 3.08 microns (aperture-masking) and 4.67 microns (direct
imaging) did not detect the companion, but 
were able to place 2-$\sigma$ upper limits on the contrast ratio of 500:1,
700:1 and 700:1 respectively. All observations reported in this
paper as well as previously published HST observations, are summarized
in Table~1, with the spectral-energy distribution shown in
Figure~\ref{figSED}. In addition to these observations, the paper of
\citet{Lopez97} provided model-fits to Infrared Spatial Interferometer
(ISI) observations with a dust
`clump' near the location of Mira~B. They reported the separation and
position angle of this clump to be 0.55$\arcsec$ and 120 degrees, but
with no estimate of uncertainty due to the limited uv coverage of the
array. The flux ratios of the clump were not given, although the dust has
an optical-depth near unity and a linear size of 17 AU.

\begin{deluxetable}{lllrrrrrr}
\tabletypesize{\scriptsize}
\tablewidth{0pt}
\tablecaption{Observation summary for Mira~A+B, with results of binary
  model fits.}
\tablecolumns{9}
\tablehead{\colhead{Wavelength}  & \colhead{Instrument}
& \colhead{JD} & \colhead{Pulsation} & \colhead{Contrast B/A}
& \colhead{Sep}& \colhead{PA}& \colhead{Mira~AB Flux\tablenotemark{b}}& \colhead{Mira~B Flux\tablenotemark{b}} \\ 
\colhead{($\mu$m)} & & & \colhead{Phase} & & \colhead{(mas)} & \colhead{(degs)} & \colhead{(Mag[M] or Jy[J])} & \colhead{(Mag[M] or Jy[J])}
}
\startdata
0.346 & FOC & 2450063.3 & 0.70 &  11.0$\pm$1.6      & 578$\pm$2\tablenotemark{a}  & 108.3$\pm$0.1\tablenotemark{a} & 10.14M & 10.23M \\ 
0.501 & FOC & 2450063.3 & 0.70 &  0.88$\pm$0.12     & - & - &  10.48M & 11.30M \\ 
0.374 & STIS & 2453051.8 & 0.65 &  0.80$\pm$0.04     & 546$\pm$3  & 105.8$\pm$0.3 & 11.07M & 11.95M  \\ 
0.97  & NICMOS & 2451037.3 & 0.62 & $<$0.002 & - & - & - & - \\
3.08  & NIRC  & 2453252   & 0.24 & $<$0.0014 & - & - & - & - \\
4.67  & NIRC2 & 2453909.5 & 0.22 & $<$0.0014 & - & - & - & - \\ 
18.75 & LWS & 2451423.6 & 0.77 &  0.102$\pm$0.022 & 476$\pm$20 & 99$\pm$2      & 1050J & 107$\pm$28J \\
17.65 & LWS & 2452506.6 & 0.02 &  0.11$\pm$0.03   & 511$\pm$34 & 96$\pm$4      & 1020J & 112$\pm$35J \\ 
10.7  & LWS & 2453248   & 0.24 &  0.022$\pm$0.003 & 470$\pm$14 & 102.7$\pm$1.6 & 4190J & 92J \\ 
12.5  & LWS & 2453248   & 0.24 &  0.021$\pm$0.003 & 489$\pm$14 & 105.2$\pm$2.0 & 2450J & 51J \\ 
7.9   & T-ReCS & 2454053.6 & 0.65 &      $<  $0.005 & -          & -         & 1780J & - \\
9.7   & T-ReCS & 2454053.6 & 0.65 & 0.025$\pm$0.012 & 480$\pm$30 & 101$\pm$3 & 2040J & 51J \\ 
12.3  & T-ReCS & 2454053.6 & 0.65 & 0.027$\pm$0.009 & 467$\pm$30 & 104$\pm$3 & 1410J & 38$\pm$13J \\ 
18.3  & T-ReCS & 2454055.6 & 0.65 & 0.065$\pm$0.02  & 487$\pm$30 & 105$\pm$3 & 1690J & 110$\pm$35J \\ 
\\
\enddata
\tablenotetext{a}{Binary fit for JD 2450063.3 comes from an average of several measurements between
0.278 and 0.501\,$\mu$m, as described in \citet{Karovska97}.}
\tablenotetext{b}{Flux error are 15\% or 0.15 mag unless indicated otherwise. The magnitude scale is in Vega magnitudes.}
\label{tabBinaryFits}
\end{deluxetable}

\section{Modeling the IR emission}
\label{sectModelling}

In order to find a preferred model for the mid-infrared emission
coming from the region around Mira~B, we must explain the
time-dependent astrometry and the total flux (i.e. the energetics) of
the emission. We will now examine several hypotheses in turn.

\subsection{A Dust Clump in the Wind?}
\label{sectClump}

If the radiation near Mira~B were re-processed radiation from
  Mira~A, then the flux would change with the Mira variability cycle.
There is a decrease in flux at the most temperature-sensitive wavelength
of 10\,$\mu$m for the feature near Mira~B occurring 
over the interval between the Keck (LWS) and Gemini (T-ReCS)
observations. This decrease co-incides with a decrease in the flux from
Mira~A due to the Mira variability cycle (see
Table~\ref{tabBinaryFits}), consistent with the re-processsed
  radiation hypothesis.
This is further supported by the modeling of \citet{Lopez97}, where
a constant amount of dust was used to model observations at several
epochs. One possible origin
is therefore re-radiated flux from an unusually dense clump in the wind from Mira~A. 


The likelihood of a clump of dust in the wind causing the mid-infrared
emission is not great, as an optical-depth of order
unity is required to produce enough emission. In turn, this requires a
wind over-density of at least 25 (based on full
silicate condensation at solar metallicity).

We can be slightly more sophisticated and model this dust clump as a uniform slab, assuming
that the dust is made from a combination of fully-condensed corundum
and olivine at solar metallicity. We take the optical constants for olivine from
\citet{Dorschner95} and those of corundum from \citet{Koike95} and
calculate dust opacities as in \citet{Ireland06}. From three-color
photometry, we can derive a column density, a temperature and an
angular extent as the radius of an equivalent uniformly-illuminated
disk. For the near-maximum LWS data, we obtain
0.28\,g\,cm$^{-2}$, 310\,K and a disk radius of 59\,mas, and 
for the near-minimum T-ReCS data we obtain 0.2\,g\,cm$^{-2}$, 240\,K and
96\,mas. These temperatures are also consistent with the
assumption of heating of the dust by Mira~A for moderate (roughly 45
degree) angles between the clump to the line-of-sight. These column
densities can be compared to those of a 100\,AU deep column in the
time-averaged wind at the $\sim$50\,AU projected separation, which
totals only 0.007\,g\,cm$^{-2}$.

These calculations are somewhat dependent on the assumed and unknown
dust opacities, but they serve to provide plausible physical model parameters for
the emission. The large required column densities are due to the emission in the
12\,$\mu$m bandpasses being brighter than expected from optically-thin
emission in the 10 and 18\,$\mu$m bands. Alternatively, one can think
of the emission at 10 and 18\,$\mu$m as being suppressed due to the dust
being optically-thick. 

The modeling of the Fourier data in \citet{Lopez97} put a dust `clump' near the location of
Mira~B between 1990 and 1994. The position angle of 120 degrees and separation of
0.55\,$\arcsec$ places this emission inside the early-1990s position
of Mira~B but outside the position of the dust clump reported in this
paper. Therefore, not only is this clump of dust inexplicably over-dense,
but its apparent motion is in the opposite direction to the outflow
from Mira~A, which should move outwards at 14\,mas per year according to the
outflow measurements of \citet{Knapp98} with an assumed distance of
107\,pc. This would mean that if the dust `clump' seen by Keck and
Gemini moved with the Mira~A outflow, it would have been at
$\sim0.31\arcsec$ during the \citet{Lopez97} epoch. Despite the
simple modeling in that paper, it is not plausible that the clump
separation was overestimated by a factor of 1.8, or that in addition
to the 0.55$\arcsec$ clump, a (brighter) $\sim$0.31$\arcsec$ clump was missed in
the analysis.

Therefore, due to both the anomalously high optical depth of
dust near Mira~B and the motion of the emission counter to 
to the prevailing wind from Mira~A, we conclude that the mid-infrared
emission is inconsistent with a clump in the wind from
Mira~A.

\subsection{Emission from the Wind Interaction Region?}

Although the variability of the mid-infrared flux near Mira~B is
  consistent with re-processed radiation from Mira~B, the phase
  coverage and signal-to-noise of the data in
  Table~\ref{tabBinaryFits} is clearly inadequate to exclude processes
  independent of Mira~A's radiation field. Therefore, we will check
  if the energy involved in the wind interaction region is enough to
  explain the 
mid-infrared emission. \citet{Wood06} model the early 2004 outflow from Mira~B as
a 450\,km\,s$^{-1}$ wind losing $2.5\times10^{-13}
M_{\sun}$yr$^{-1}$. This wind is variable, with evidence that it was
weaker in the late 1990s and stronger in the IUE era \citep{Wood02}. As this wind
encounters the more dense wind from Mira~A, it will lose it's energy
in a shock front. This shock luminosity ($\frac{1}{2}\dot{M}v^2$) will
be $2.6\times10^{-6} L_\sun$, which is a tiny fraction of the
mid-infrared luminosity coming from near Mira~B, which is $\sim 10 L_\sun$.

Thus, if the mid-infrared emission were to originate from the
wind-interaction region, it would have to be due to a quasi-static
increase in the density of Mira~A's wind due to the wind from
Mira~B. According to the model of Section~\ref{sectClump}, this region
has to have at least $3.8 \times 10^{-6}$ solar-masses of material,
over a length scale of $10-20$\,AU. 20\,AU is roughly the
maximum distance over which the flux would still be largely unresolved
in our images, once projection effects are taken into account. 


The simplest geometry in which to consider the outflow from Mira~B is
one where the wind from Mira~B is largely perpendicular to the wind from
Mira~A.  In this geometry, the mass excess due to the slowing of
Mira~A's wind by Mira~B's wind is of order $\dot{M_B}D/v_w$ over
length scales $D$, which is $\sim10^{-12} M_\sun$, or 6 orders
of magnitude too little material.

The outflow from Mira~B may be nearly spherically-symmetric, including
a component as just described, as well as a region where the two winds
collide nearly head-on, possibly forming a stagnation
point. Calculation of the stagnation point position is a simple
balance of dynamic (ram) pressures $\rho_w v_w^2$, where
$\rho_w$ is the wind density and $v_w$ the wind speed (for either
Mira~A or Mira~B). This stagnation
point is significantly influenced by the gravitational field from
Mira~B: both by an change in density $\rho_w$ of Mira~A's wind and
an increase in velocity. 

We have taken these effects into account using Newtonian physics, and
calculated the stagnation point position. With Mira~B wind
parameters of $2.5\times10^{-12}$\,$M_\sun/yr$ at 450\,km\,s$^{-1}$, Mira~A
wind parameters of $4.4\times10^{-7}$\,$M_\sun/yr$ at
6.7\,km\,s$^{-1}$ \citep{Knapp98} a Mira~AB distance of 90\,AU and a
Mira~B mass of 0.7\,M$_{\sun}$ (see Section~\ref{sectDisk}), this
stagnation point occurs only 1\,AU from Mira~B (projection effects not
taken into account). An upper limit for this distance is the 3.7\,AU
calculated in \citet{Wood02} for the stronger IUE era wind, with
different assumed wind and orbital parameters. 
Therefore, a large
build-up of dust at this stagnation point could not form a region
geometrically large enough to provide the mid-infrared flux.



\subsection{Emission from a Centrally-Illuminated Accretion Disk}
 
T~Tauri stars are well known to have significant mid-infrared
excesses. As accretion onto Mira~B occurs through a disk, we can
also ask if the emission is consistent with expectations from known
and well-characterized accretion disks. By comparing the emission of the Mira~B system
to T~Tauri stars (i.e. systems with significant accretion luminosity)
as in Figure~\ref{figSED}, it is clear that although the blue and
ultraviolet accretion luminosity is typical, the very large
mid-infrared flux from Mira~B is anomalous by more than a factor of
50. Therefore, this emission must be due to a mechanism atypical of 
isolated systems with a disk and outflow.
Furthermore, the problem of the in-phase variation with the Mira~A primary
presents an additional stumbling block to this model.

\subsection{The Edge of a Self-Shadowed Accretion Disk}
\label{sectDisk}

\begin{figure}
 \plotone{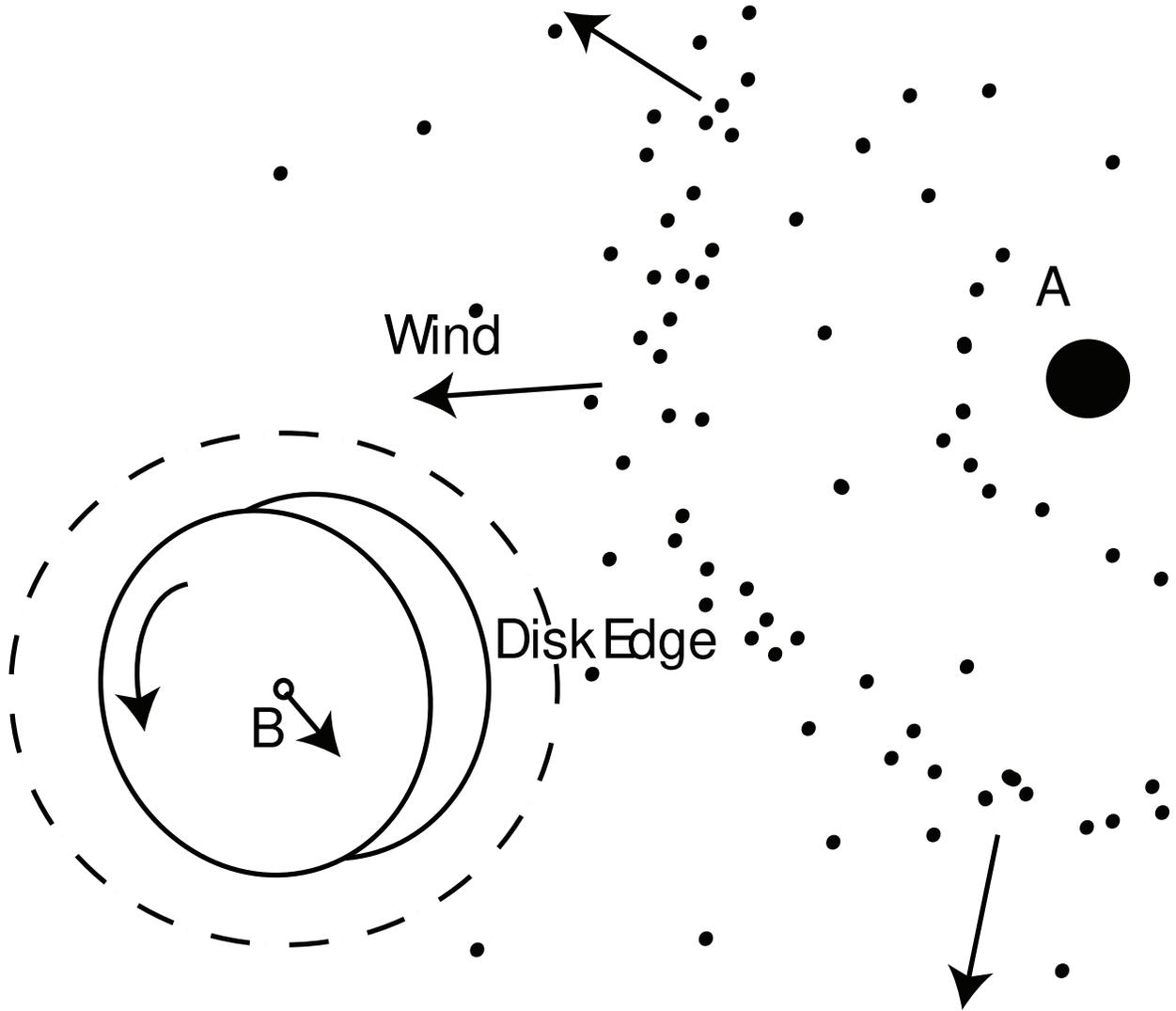}
 \caption{A diagram showing the disk model geometry for Mira~AB. The dashed line represents the Bondi-Hoyle
 accretion radius.}
\label{figQuickDiagram}
\end{figure}

We will now examine a model of the mid-infrared emission near Mira~B
that is caused by illumination of an accretion disk around Mira~B by
radiation from Mira~A. In this section we will not discuss the
possibilities of this disk being around a white dwarf or a main
sequence star. This disk will be supposed to be in the plane
of the Mira~AB orbit, so that one edge is illuminated and the
opposite edge is in shadow. This co-planar geometry is expected if
angular momentum is preferentially accreted in the plane of the
orbit. It is illustrated in Figure~\ref{figQuickDiagram}. First, we shall review the theory of wind 
accretion as it applies to Mira~AB.

The accretion of the wind from Mira~A onto Mira~B is given by the
formulae of Bondi-Hoyle-Lyttleton accretion
\citep[e.g.][]{Edgar04}. In its simplest form, 
the wind is gravitationally focused into a wake
behind the accretor, and the wake is accreted wherever it is
gravitationally bound to the accretor. This gives an accretion radius,
which is the radius of a circular area that corresponds to the effective
cross-section of the accretor: 

\begin{equation}
  R_{\rm acc} = \frac{2 G M_B}
      {v_w^2 + v_{\rm orb}^2}.
 \label{eqnBHRadius}
\end{equation}

Here $M_B$ is the accretor (i.e. Mira~B) mass, $v_w$ the wind velocity
and $v_{\rm orb}$ the orbital velocity. This formula is most
applicable to highly supersonic flows, which is the case for Mira~B
accreting the wind from Mira~A. This accretion radius can be used to
find an accretion rate:

\begin{equation}
 \dot{M}_{\rm acc} = \frac{G^2 M_B^2 \dot{M}_A}
      {r^2_{\rm orb}v_w(v_w^2 + v_{\rm orb}^2)^{3/2}}.
 \label{eqnBHAcc}
\end{equation}

The accretion luminosity is then given by the simple formula $L=G
\dot{M}_{\rm acc} M_B/r_B$, where $r_B$ is the radius of the
companion. This formula assumes an equilibrium situation, which may
not be the case if material is accumulating within the
disk. 


In order to test hypotheses that the accretion disk around Mira~B has been 
directly observed, we should first examine some basic parameters of the
system. The best estimate for $\dot{M}_A$ is $4.4 \times 10^{-7}
M_\odot$ per year, with $v_w=6.7$\,km\,s$^{-1}$ \citep{Knapp98}. This mass-loss
rate is based on a CO to H$_2$ ratio of $5 \times 10^{-4}$ which is in turn
based on a near-solar metallicity. If Mira~A has sub-solar
metallicity as expected from its kinematic association with the
thick-disk population, then this mass-loss rate will be
underestimated. The orbit of Mira~B is poorly known due to it's long period, but if we assume
a circular orbit and fix the total system mass to
1.5\,$M_\sun$, then the astrometry of \citet{Prieur02} combined with
the most recent HST epoch (in Table~\ref{tabBinaryFits}) gives a crude
orbital radius of $r_{\rm orb} = 90$\, AU, an inclination is 63
degrees and a period of 610 years. Changing the assumed total system
mass to 2.0\,$M_\sun$ only changes the orbital radius to 85\, AU and
the inclination to 66 degrees. For the total mass 1.5\,$M_\sun$ orbit,
we have an orbital velocity $v_{\rm orb}=3.2$\,km\,s$^{-1}$ and an accretion
radius in AU of 32\,$M_B/M_{\sun}$ according to
Equation~\ref{eqnBHRadius}. 


The maximum size of the accretion disk around Mira~B will be approximately the
Bondi-Hoyle accretion radius, as beyond this radius the wind from Mira~A will have enough
energy to both escape from the gravity of Mira~B and transfer some
momentum onto material orbiting around Mira~B. Conversely, the minimum accretion
radius is the size of the disk measured from the difference in
astrometry between Mira~B and the mid-infrared emission. In the plane
of the sky, the mid-infrared emission is separated from Mira~B by
$6.5\pm1.4$\,AU. The error in this astrometric difference is difficult
to quantify in individual images, particularly given the possibility
that asymmetries in the atmosphere of Mira~A could move its centroid
around by of order 1\,AU. However, even if we do not count the \citet{Lopez97}
result as a single measurement, then we have 6 independent measurements of the
astrometric difference in Table~\ref{tabBinaryFits} that give the same
sign and a similar magnitude. Determination of the average separation
between Mira~B (from the UV) and the mid-infrared emission therefore
appears statistically robust with reliable errors. The true separation
of Mira~B and the mid-infrared emisson is then 
$10.6 \pm 2$\,AU, given that the vector between Mira~A and B is at 38 degrees to the
line-of-sight from our preferred orbit. This is consistent with a disk $\sim13$ AU in size once the
effect of the curved geometry of the disk edge is taken into account.

This disk size is is consistent with Bondi-Hoyle accretion theory,
predicting an accretion radius range from 16 to 22\,AU for $M_B$
between 0.5 and 0.7 $M_{\sun}$. Importantly, the disk size shows that
the accretion radius cannot be under-predicted using Bondi-Hoyle
accretion theory by more than a factor of $\sim$2.

We must also establish that the amount of mid-infrared emission seen
from the edge of the disk around Mira~B is consistent with the
observed emission. Already in Section~\ref{sectClump} we have seen
that the emission is consistent with a region of cross-sectional area
about 100 square AU at temperature 310\,K (from the slab model with a
radius of 59\,mas). If Mira~B is on the far side of Mira~A as viewed 
from Earth, we see the edge of the
disk that is directly illuminated by Mira~A. 

If the disk edge has a temperature
of 310\,K, then its hydrostatic equilibrium density distribution is
Gaussian with Full-Width Half-Maximum (FWHM) of 1.5\,AU at a 10\,AU
radius. This disk of diameter
20\,AU would only give a 100 square AU cross-sectional area if the
disk remained optically-thick to radiation from Mira~A out to several
times the density FWHM. However, given the significant uncertainties
and simplifications in the slab model (including the opacities
themselves), this emission is certainly consistent with an origin in a
side-illuminated accretion disk. 

\begin{figure}
 \plotone{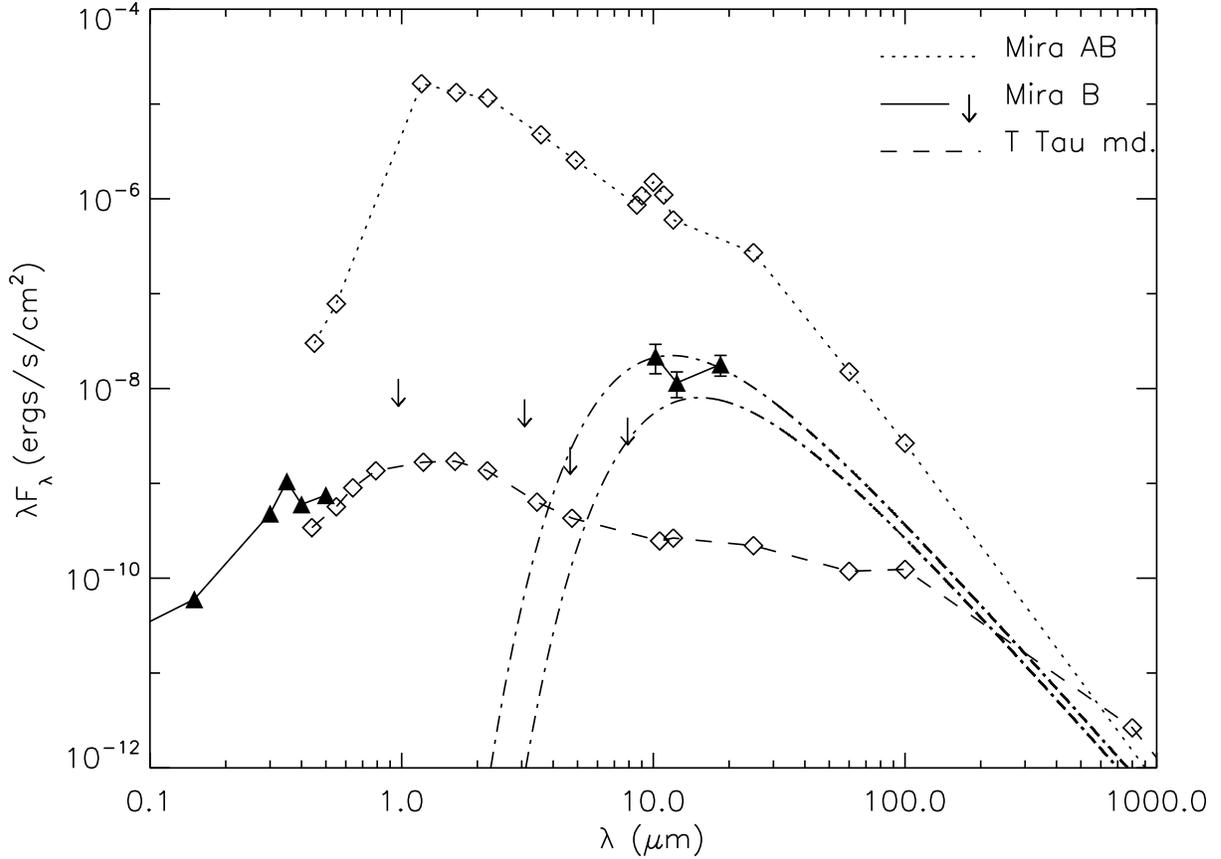}
 \caption{SED for Mira~B, the median Mira~AB spectrum, and the median
 SED for T Tauri stars in Taurus placed at the same distance as Mira
 AB. Upper limits are 2-$\sigma$. The mid-infrared measurements from
 Keck and Gemini are averaged to represent a median flux between
 pulsation maximum and minimum, and the upper limit contrast ratios
 are also referenced to the median Mira~A flux. The dot-dashed lines
 are representative black-body curves at the `slab' model temperatures of
 240 and 310\,K.}
\label{figSED}
\end{figure}

Note also that the spectral energy
distribution shown in Figure~\ref{figSED} is not consistent with the
optical depth of unity temperature being the same throughout the
mid-infrared: the sillicate emission peak and the upper limit at
7.9\,$\mu$m (from the near-minimum Gemini epoch) demonstrate that the outer surface of
the disk has to be heated. This is an essential part of any physical
model of the disk, where radiative energy flows inwards.




Given that we have established that Mira~B has a $\sim$10\,AU radius
dusty disk around it, we should establish whether the term
`protoplanetary' can be applied to it. This term implies that there is
or will be enough material in the disk to form planets. According to
the prescription of \citet{Hayashi81}, the minimum-mass solar nebula
had 2 Jupiter masses out to a radius of 10\,AU. The total mass
accreted onto Mira~B during the lifetime of Mira~A is of order 7
Jupiter-masses, assuming that Mira~B has 0.7 solar-masses and Mira~A
loses 0.6 solar masses while on the Asymptotic Giant Branch. The
ability for such a disk to form planets clearly depends on a number of
factors, such as the role of viscosity in the disk and the maximum
accretion rate that will occur as Mira~A's wind transitions to a
radiatively driven wind with a $\sim$20 times increase in mass-loss
rate. 

According to the viscous evolution prescription of
\citet{Alexander06} with their `typical' disk parameters, the current
equilibrium disk mass around Mira~B is only 0.26\,$M_J$ based
on a $M_B=$0.7\,$M_\sun$, an accretion rate given by
Equation~\ref{eqnBHAcc} and a disk radius of 10\,AU. The time scale
for viscous evolution of the disk is $\sim3 \times 10^4$ years. As the
time scale for Mira~A's evolution is based on the time between thermal
pulses which is of order $10^5$ years \citep{Vassiliadis93}, the disk
will likely maintain a near-equilibrium mass. At the time Mira~A
becomes a white dwarf, and the mass-loss rate from Mira~A is
$\sim10^{-5}$ solar-masses per year, the equilibium disk mass will
have increased to several Jupiter-masses, certainly enough to
justify the title `protoplanetary disk'.

\section{Discussion}
\label{sectDiscussion}

\subsection{Nature of Mira~B}

Now that we have established that Mira~B has a $\sim10$\,AU accretion
disk around it, we should re-examine the idea that Mira~B is a
main-sequence star and not a white dwarf as often identified in the literature.
The Mira~AB system is quite different to symbiotics such as MWC~560, where
the accretion luminosity is 1000 instead of
0.1\,$L_\sun$. However, accretion luminosity should vary by several
orders of magnitude for symbiotics due to differing accretion rates,
and as Mira~AB is unique in its relatively well-known geometry and
wind properties, it is difficult to make quantitative comparisons.
A second general property of symbiotics such as MWC~560 is that there
are spectral features with wind absorption and emission profiles with
widths that are one to several thousand km\,s$^{-1}$ \citep{Schmid01}, comparable to or greater
than the escape velocity from the surface of a white dwarf. Despite
the large number of lines observed for Mira~B \citep[e.g.][]{Wood06},
there are no lines with half-widths of several thousand km\,s$^{-1}$ as one
would expect if Mira~B were a white dwarf.

The analysis of Section~\ref{sectDisk} demonstrated that the observed size of
the accretion disk around Mira~B is consistent with Bondi-Hoyle
accretion, and that this accretion theory cannot underestimate the
accretion rate onto Mira~B by more than a factor of a few. 
On the other hand if we assume that Mira~B is a 0.6\,$M_\sun$ white dwarf, then the same
analysis gives an accretion rate of $5\times10^{-9} M_{\sun}$/yr and an 
accretion luminosity of
9.4L$_\sun$. This is inconsistent with the measured accretion luminosity
of Mira~B, which is $\sim 0.1-0.5 L_{\sun}$, depending on what
fraction of
the flux is from accretion and what the extinction is. The value of
$0.5 L_{\sun}$ is an upper limit based on all the measured luminosity
at $\lambda<0.5 \mu$m being due to accretion and $A_U=1$, the extinction
calculated from full silicate condensation in Mira~A's wind. There is
a possibility that the accretion luminosity is lower than expected for
Mira~B due to a non-equilibrium situation where mass is accreted into
a disk but not onto Mira~B itself. The disk evolution time-scale
discussed in Section~\ref{sectDisk} makes this a highly unlikely
scenario. If on the other hand, we assume that Mira~B is a 0.7\,$M_\sun$
main-sequence star, then the accretion luminosity is 0.2$L_{\sun}$,
completely consistent with the measured luminosity of Mira~B.

Finally, if Mira~B were a main-sequence star, we could expect to see
spectral features that indicate the presence of a main-sequence star,
although these features could be very weak. The weak features are due to
significant veiling by the accretion flux: a good example of this is
for T~Tauri stars can be found in the spectrum of BP~Tau, which is
significantly veiled for all wavelengths shorter than 0.5\,$\mu$m
\citep{Gullbring00}. The Faint Object Camera (FOC) spectra partially
published in \citet{Karovska97} do indeed show weak (i.e. veiled)
spectral features of an early-mid K dwarf. 

\begin{figure}
 \plotone{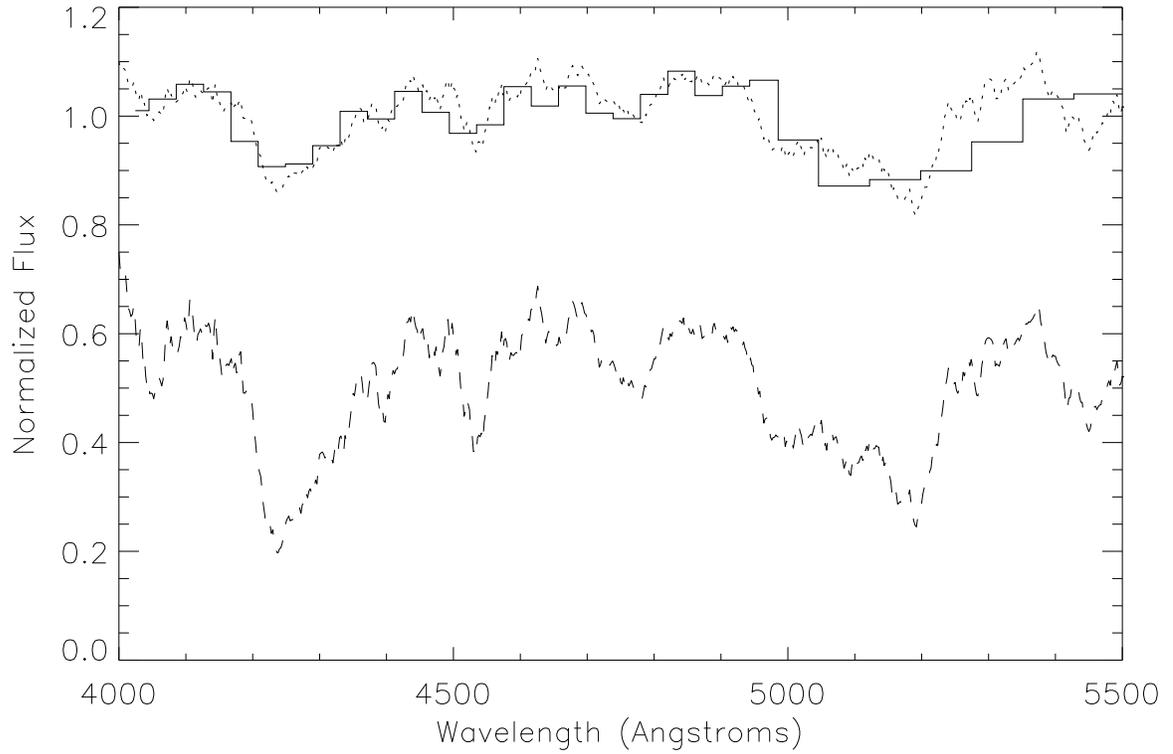}
 \caption{The spectrum of Mira~B from the HST Faint-Object Camera (solid line)
 taken on Dec 11, 1995. Signal-to-noise is $\sim$20. Overlaid (dashed line) is a
 model spectra from the NextGen model series \citep{Hauschildt99} at
 4000\,K and log(g)=5.0, and the same spectrum veiled by 80\% at
 4000\,\AA and 60\% at 5500\,\AA.}
\label{figKDwarf}
\end{figure}

We have examined the full
wavelength range of this spectrum (including that not published),
derived from HST data set X31G010AT, and displayed the long-wavelength end of
it in Figure~\ref{figKDwarf}. We took the wavelength scale calibration
from Chapter~8 of the FOC manual, referencing the zero point to the Mg
II doublet near 2800\,\AA prominent in the Mira~B
spectrum. Prominent spectral features at 4250\,\AA and 5100\,\AA
are both detected at $>4\sigma$, matching a veiled model $\sim$4000\,K
dwarf spectrum well. The $\sim$60\% veiling in V-band places the
photospheric spectrum of Mira~B at an absolute V magnitude of
$7.2\pm0.2$, also consistent with a 0.7\,$M_\sun$ K5 dwarf (with
$\sim<0.4$ mag of extinction at V band).

All remaining possibilities for the nature of Mira~B that could
influence our conclusions consist of some kind of
exotic geometries. An example could be a triple system with a radial velocity
signature less than $\sim 20$km\,s$^{-1}$ (so that it is not betrayed by 
ultraviolet spectral shifts) and a close-system separation less than
$\sim 10$\,AU. In the absence of any data to support such a
possibility, we will eliminate these possibilities with an appeal to Occam's razor. 
Therefore, we conclude that Mira~B is indeed a
main-sequence star, with a most probable mass of roughly 0.7\,$M_\sun$.

\subsection{How common are systems like Mira~B?}

From the standpoint of population synthesis, we can examine how common
we expect Mira~B like systems to be. Our current galactic
location in-between spiral arms means that the column-integrated star
formation rate is about a quarter of what it was when Mira~A's
progenitor was born \citep{Schroder03}.
In order to accrete at least a Jupiter mass via wind-accretion, a
wind-accreting binary should have log(P) between 4 and $\sim$6.5 in
days (the upper limit for P is dependent on primary and secondary
mass). According to \citet{Duquennoy91}, $\sim$23\%
of solar-mass stars have companions with log(P) in this range.
These results, taken together, imply that the birth-rate of Mira~B-like
disks around companions to 1-2\,$M_{\sun}$ stars should be similar to the
local birth-rate of 1-2\,$M_{\sun}$ stars themselves.

This immediately raises the question: where are the disks that formed
by wind-accretion? Most debris disks in the solar neighborhood are
around stars likely belonging to young associations
\citep{Moor06}. However, these debris disks are also generally cool
and large, while a wind-accretion disk will be truncated by
the dynamical pressure of the wind at roughly the Bondi-Hoyle
accretion radius. The disks around secondaries in
wind-accretion binaries should grow viscously after the primary dwindles into
a white-dwarf, but clearly this process must leave debris disks much
less often than the large primordial disks around young stars.

The expected abundance of disks around companions to young white
dwarfs is a clearly testable hypothesis of the Bondi-Hoyle truncation
model of wind-accretion disks presented here. However, in order to
test it, a sample of very young white dwarfs in binaries would have to
be chosen. There are some such systems discovered through UV excesses: 
\citet{Burleigh97} discusses many of them and some of the difficulties
in forming a  near-complete sample. Other systems are discovered
through infrared excesses to white dwarfs: many of these have been
imaged with HST \citep{Farihi06}. Disk fraction around the main
sequence companions should approach unity for appropriate separations
and ages less than several times the disk viscous evolution time-scale,
i.e. a few $\times 10^5$ years.


\section{Conclusions}

This paper has presented evidence that Mira~B is a $\sim$0.7\,$M_\sun$
main sequence star surrounded by a $\sim10$\,AU radius disk. The
new observations on which this is based were made possible by the
high-contrast imaging capability of non-redundant interferometry on
single-aperture telescopes. Mira~AB is unique amongst wind accretion
binaries in its well known binary and wind parameters, but this class
of system is predicted to be relatively common as the end product of
roughly 1 in 5 star systems with a solar-type primary. Systems like
Mira~AB should produce the clear observational signatures of an
accretion disk around the secondary for at least a few $\times 10^5$
years after the primary becomes a white dwarf.

\acknowledgements

We gratefully acknowledge the support of Charles Townes, the
assistance of Marc Kassis in making the Keck observations, the
assistance of Adwin Boogert in planning the Gemini observations and helpful
discussions with Nuria Calvet and Klaus Pontoppidan. M.I. would like to acknowledge
Michelson Fellowship support from the Michelson Science Center and the
NASA Navigator Program. This research has made use of the SIMBAD
database, the INES principle center and the IRAF point source
catalog. J.D.M. acknowledges support from the grant NASA-JPL
1267021. Some of the data presented herein were obtained at the
W.M. Keck Observatory, which is operated as a scientific partnership
among the California Institute of Technology, the University of
California and the National Aeronautics and Space Administration. The
Observatory was made possible by the generous financial support of the
W.M. Keck Foundation. The authors wish to recognize and acknowledge
the very significant cultural role and reverence that the summit of
Mauna Kea has always had within the indigenous Hawaiian community.  We
are most fortunate to have the opportunity to conduct observations
from this mountain. Based in part on observations obtained at the
Gemini Observatory, which is operated by the Association of
Universities for Research in Astronomy, Inc., under a cooperative
agreement with the NSF on behalf of the Gemini partnership: the
National Science Foundation (United States), the Particle Physics and
Astronomy Research Council (United Kingdom), the National Research
Council (Canada), CONICYT (Chile), the Australian Research Council 
(Australia), CNPq (Brazil) and CONICET (Argentina).


\end{document}